\documentclass[final,3p,times]{elsarticle}

\usepackage{amssymb}

\usepackage{booktabs}
\usepackage{caption}
\usepackage{graphicx}
\journal{Physica A}
\usepackage[font=small,labelfont=bf,labelsep=none]{caption}
\usepackage[ruled,linesnumbered]{algorithm2e}
\usepackage[section]{placeins}
\usepackage{enumitem}

\usepackage{comment}

\usepackage[colorlinks, bookmarksopen, bookmarksnumbered, citecolor=blue, linkcolor=blue, urlcolor=black]{hyperref}

\begin{document}

\begin{frontmatter}



\title{HGC: A hybrid method combining gravity model and cycle structure for identifying influential spreaders in complex networks}


\author[1]{Jiaxun Li}
\author[1]{Yonghou He}
\author[1]{Zhefan Dong}
\author[1]{Li Tao \corref{cor1}}
\ead{tli@swu.edu.cn }

\address[1]{School of Computer and Information Science, Southwest University, Chongqing 400715, China}

\cortext[cor1]{Corresponding author}


\begin{abstract}
Identifying influential spreaders in complex networks is a critical challenge in network science, with broad applications in disease control, information dissemination, and influence analysis in social networks.
The gravity model,  a distinctive approach for identifying influential spreaders, has attracted significant attention due to its ability to integrate node influence and the distance between nodes.
However, the law of gravity is symmetric, whereas the influence between different nodes is asymmetric.
Existing gravity model-based methods commonly rely on the topological distance as a metric to measure the distance between nodes. Such reliance neglects the strength or frequency of connections between nodes, resulting in symmetric influence values between node pairs, which ultimately leads to an inaccurate assessment of node influence.
Moreover, these methods often overlook cycle structures within networks, which provide redundant pathways for nodes and contribute significantly to the overall connectivity and stability of the network.
In this paper, we propose a hybrid method called HGC, which integrates the gravity model with effective distance and incorporates cycle structure to address the issues above.
Effective distance, derived from probabilities, measures the distance between a source node and others by considering its connectivity, providing a more accurate reflection of actual relationships between nodes.
To evaluate the accuracy and effectiveness of the proposed method, we conducted several experiments on eight real-world networks based on the Susceptible-Infected-Recovered model. The results demonstrate that HGC outperforms seven compared methods in accurately identifying influential nodes.

\end{abstract}




\end{frontmatter}


\section{Introduction}
\label{}

 One of the most significant challenges in network science is identifying influential spreaders, commonly referred to as important nodes. These nodes play a critical role in the propagation of information \cite{hosni2020minimizing} \cite{xu2019identifying} and diseases \cite{zhu2018analysis} \cite{yaomodeling} across the network. For example, during an infectious disease outbreak, immunizing key individuals can help prevent a large-scale epidemic \cite{chaharborj2022controlling}. Effectively identifying these nodes can lead to better strategies for controlling epidemics, enhancing information dissemination, and maximizing influence in social networks.

Numerous methods have been proposed to identify important nodes in complex networks. Based on the network topology information they rely on, these methods can be broadly categorized into two types: methods based on global information, and methods based on local information.
Methods based on global information consider the structural information of the entire network. Betweenness centrality \cite{bc} (BC) and closeness centrality \cite{cc} (CC) are two typical methods of this type. BC quantifies the frequency with which a node appears on the shortest paths between pairs of nodes in the network. CC measures how close a node is to all other nodes, typically by calculating the sum of the shortest paths between the node and all others in the network. These methods can consider node information in correlation with the network's topology, but they have high computational complexity and are unsuitable for large-scale networks. 
Additionally, cycles in networks play significant roles in both structural organization and functional implementation. 
Building on this foundation, Fan et al. \cite{cr} considered the cycle structure in networks and defined the cycle number matrix and cycle ratio to quantify the importance of nodes. However, this method is not effective when the network is sparse.

Local information methods, such as degree centrality \cite{DC} (DC) and the H-index \cite{h-index}, provide more computational efficiency compared to global methods, making them suitable for large-scale networks. DC posits that the more neighbours a node has, the greater its influence. However, DC only calculates the number of neighbours and does not consider the attributes of the neighbouring nodes. Therefore, Li et al. \cite{RDP} proposed a method called RDP, which can obtain higher-order neighbourhood degree information with just a few iterations and has very low computational complexity. 
However, these methods are limited in that they rely solely on the information of nodes and their neighbours, thereby ignoring the overall topological structure of the network. This may result in the identification of local optima rather than global optima.

Recently, the gravity model, as a method based on local and global information, has received widespread attention because it integrates both the distance between nodes and their influence, providing a more comprehensive approach to node identification. 
Inspired by the gravity law, Ma et al. \cite{G} proposed two models (G and G+). These models use the K-shell value of nodes as mass and the shortest path distance between two nodes as distance. 
Afterwards, Li et al. \cite{GM} introduced the gravity model (GM), using the degree value of nodes as mass, and further proposed the local gravity model (LGM), which only considers the neighbouring nodes within a truncation radius. 
Moreover, Li et al. \cite{dkgm} combined the K-shell value and the K-shell iteration factor as node mass to propose the DK-based gravity model (DKGM). Subsequently, they considered multiple characteristics of nodes and proposed the multi-characteristics gravity model \cite{mcgm} (MCGM). Zhu et al. \cite{hvgc} proposed a new gravity centrality method called HVGC, based on the H-index. 

However, these gravity-based methods assume that the influence between nodes is symmetric, assigning the same distance to any pair of adjacent nodes while
neglecting the interaction strength between them, which refers to the connectivity of the nodes. 
This simplification contradicts the principles of information propagation in real-world networks. In practice, information is distributed across different paths based on the connectivity of the nodes, so the distance between any two nodes should not always be the same. Similarly, the influence of node A on node B may not be equal to that of node B on node A. 
In addition, these methods overlook the cycle structure in networks. A cycle in the network can be defined as a closed path that starts and ends at the same node. Although the cycle is the simplest structural element in a network, it plays a crucial role in network analysis. Recent studies have revealed the significant role of cycles in network functionalities such as storage, synchronizability, and controllability \cite{storage} \cite{synchronizability} \cite{controllability}. Cycle structures are ubiquitous in networks. Without the cycle structure, networks will degrade into tree-like networks. Removing a single edge or node in such a scenario will result in the network splitting and the collapse of its functions.

To address these limitations, we propose a hybrid method combining the gravity model and cycle structure to identify important nodes in networks, referred to as the HGC, which considers the influence of neighbouring nodes, the effective distance between nodes, and the cycle structure information of nodes within the network. Moreover, we incorporate structural hole theory \cite{structuralhole}, which identifies nodes that occupy advantageous positions between non-redundant contacts, to identify those that are not centrally located but still hold significant influence.
Based on our proposed method, we conducted several experiments utilizing the Susceptible-Infected-Recovered (SIR) model \cite{sir} on eight real-world networks and compared the results with seven methods. 
We used Kendall’s Tau \cite{kendall} and Jaccard similarity coefficient \cite{jaccard} to evaluate the accuracy of methods in identifying important nodes, simulated
the spreading ability of nodes with the SIR model, and utilized Monotonicity \cite{mono} to compare the distinguishing ability of methods.
The experimental results demonstrated that our method obtained more accurate ranking results, identified nodes with strong spreading capabilities, and reduced the frequency of occurrence of the same ranked nodes compared to other methods.

The remaining sections of this paper are organized as follows. In Section \ref{section2}, we give a brief introduction of basic definitions. Section \ref{section3} presents the proposed HGC method. Section \ref{section4} presents the specific
experimental setup, including datasets, compared methods, SIR model, and evaluation indicators. Section \ref{section5} presents the experimental results, 
analysis, and discussion.
We conclude our study in Section \ref{section6}.

\section{Preliminaries}
\label{section2}
In this section, we formalize the problem of influential node identification and introduce basic definitions that form the foundation of the subsequent analyses.

\subsection{Formulation of research problem}
Given a simple undirected and unweighted network $G=<V, E>$, where $V$ is the set of nodes in $G$ and $E$ represents the set of edges connected by two nodes. Denote $|V|=N$ and $|E|=M$, then the network includes $N$ nodes and $M$ edges. The adjacent matrix of network G is denoted by $A = (a_{ij})_{N\times N}$, $a_{ij}=1$ if node $i$ and node $j$ are linked, and 0 otherwise. Define the influence of a node $v_i \in V$ as $I ( v_i ) $. The goal is to find a subset $S\subset V$ of the top $k$ influential nodes that maximizes a predefined network influence objective $f(S)$. This objective can be formally expressed as $\max _{S \subset V,|S|=k} f(S)$.

\subsection{Basic definitions}
\noindent\textbf{Definition 1.} (Effective distance). Effective distance \cite{effectivedistance} (ED) is a distance abstracted from probabilities, and it provides a more detailed distance metric based on the interaction strength between nodes and their neighbours. In other words, after information passes through a node, it is distributed to different paths based on the node's connections with its neighbours. Therefore, the distance between any two nodes should not always be equal. Effective distance from node $i$ to node $j$ is defined as
\begin{equation}
    ED_{j|i} = 1-log_2(P_{j|i})\quad
    P_{j|i} = \frac{a_{ij}}{k_i}
\end{equation}
where $k_i$ is the degree of node $i$, $a_{ij}$ is the element in the adjacent matrix of graph $G$.

If multiple paths are from node $i$ to node $j$, we will use the shortest path between these two nodes. The shortest path is defined as
\begin{equation}
\label{effdistance}
    ED_{j|i} = \min \{ED_{j|i}^1,ED_{j|i}^2,ED_{j|i}^3,......\}
\end{equation}

Figure. \ref{cyclenumbermatrix}(b) shows the calculation of the effective distance between nodes. Clearly, $P_{6|7} \neq P_{7|6}$ and $D_{6|7} \neq D_{7|6}$ can be seen in Figure. \ref{cyclenumbermatrix}(b). The effective distance from node $i$ to itself is 0. An important difference with Euclidean distance is that the effective distance is asymmetric.

\noindent\textbf{Definition 2.} (Structural holes). Structural holes \cite{structuralhole} refer to gaps in a network where two individuals are not directly connected but are indirectly connected through a third node. Specifically, a structural hole exists between two nodes that do not have a direct link but are both connected to a common node. Burt proposed the network constraint coefficient to quantify the influence of structural holes. This metric measures the extent to which a node occupying a structural hole is constrained within the network, represented as follows.
\begin{equation}
\label{strhole}
    c_i = \sum_j(\mu_{ij} + \sum_{q\neq i,j} \mu_{iq}\mu_{qj})^2 \quad
    \mu_{ij} = \frac{z_{ij}}{ \sum_{j\in \Gamma (i)}^{} z_{ij} }
\end{equation}
where node $q$ represents the common neighbors between nodes $i$ and $j$. $\mu_{ij}$ denotes the proportion of effort that node $i$ invests to maintain its relationship with node $j$ out of its total effort.  $\Gamma (i)$ denotes the set of neighbors of node $i$, and $z_{ij}$ = 1 if there exists an edge between nodes $i$ and $j$, otherwise $z_{ij}$ = 0.


\section{Methods}
\label{section3}

\begin{figure*}[htpb]
    \centering
    \includegraphics[width=1.0\linewidth]{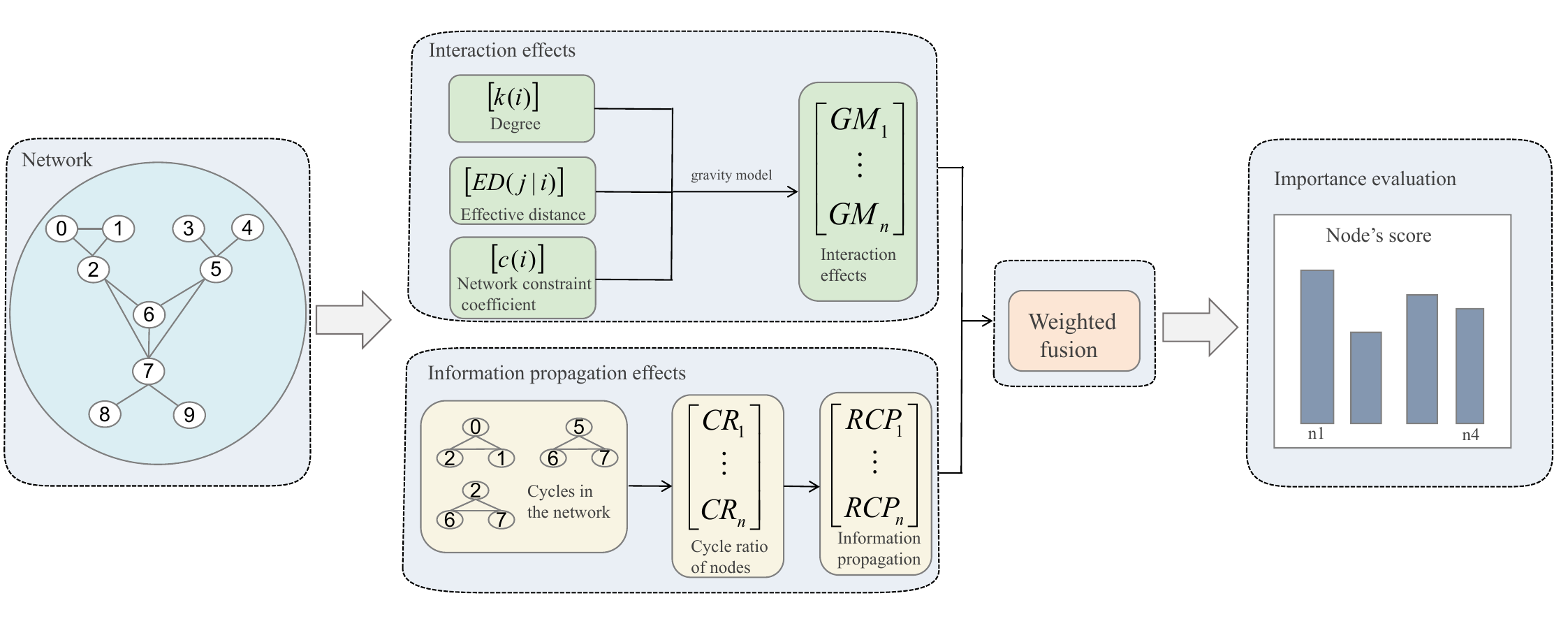}
    \caption{ HGC Overall Framework Diagram. The overall framework of HGC comprises three parts: (1) interaction effects between nodes based on the gravity model; (2) information propagation effects based on the cycle structure; and (3) weighted fusion. The framework can compute the influence scores of all nodes in the network through the three parts outlined above.
    }
    \label{framework}
    \renewcommand{\thefigure}{1}

    \end{figure*}

In the previous section, we discussed the merits and limitations of various methods for identifying important nodes in networks. 
To address these limitations, we propose the HGC method. 
The overall framework of our proposed method HGC is presented in Figure. \ref{framework}.


\subsection{Proposed method}
 The proposed method HGC consists of three components: interaction effects (see Section \ref{interaction effects} for details), information propagation effects (see Section \ref{information propagation} for details), and weighted fusion (see Section \ref{weighted fusion} for details).
 Firstly, the node degree and effective distance between nodes are incorporated into the gravity model to calculate the interaction effects, while the network constraint coefficient is used to evaluate the importance of a node's position within the network. Then, the cycle ratio is employed to quantify the significance of cycle structures in the network, with higher-order neighbour information being obtained through several iterations. Finally, a balancing factor is introduced to integrate the above two scores.
 HGC of node $i$ is defined as follows.
\begin{equation}
\label{hgm}
    HGC(i) = GM(i)+\gamma RCP(i)
\end{equation}
where $GM(i)$ represents the interaction score of node $i$ calculated by gravity model, $RCP(i)$ represents the information propagation score of node $i$, and $\gamma$ is a balancing factor.

 \begin{figure}[htpb]
    \centering
    \includegraphics[width=1.0\linewidth]{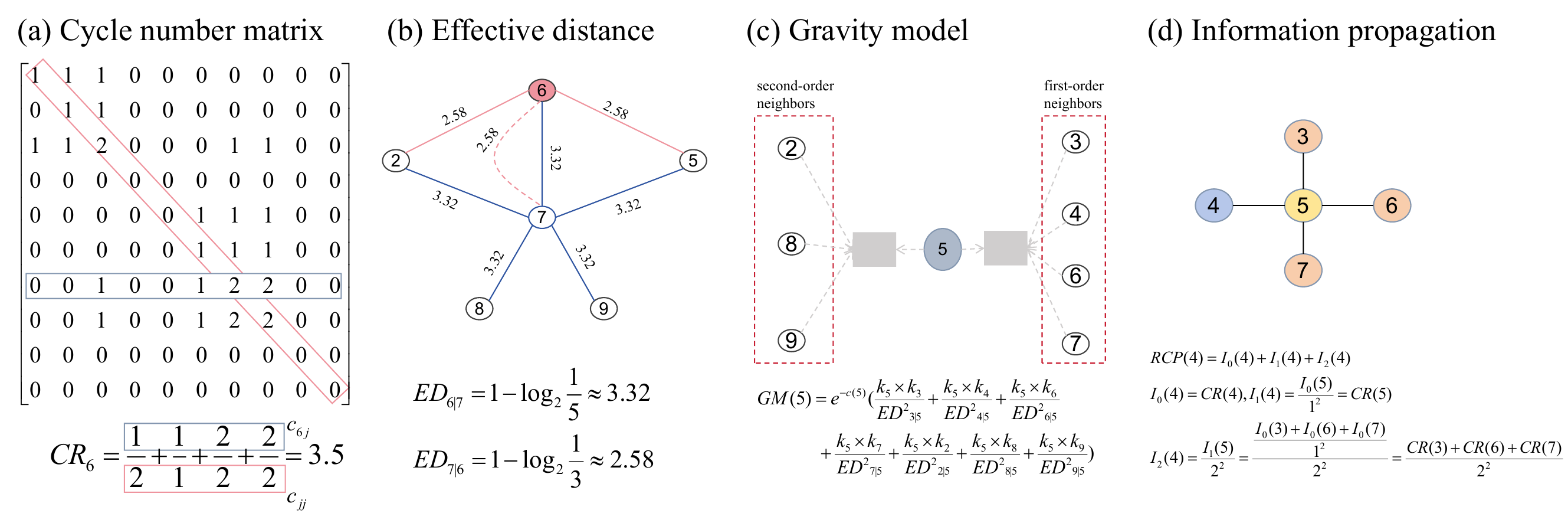}
    \caption{ Figure. (a) shows the cycle number matrix of the example network in Figure. \ref{framework} and how to calculate the cycle ratio of node 6. 
    Figure. (b) illustrates the process of calculating the effective distance between nodes.
    Figure. (c) depicts how the gravity model aggregates neighbour information.
     Figure. (d) presents how nodes iteratively collect information from their neighbours.
    }
    \label{cyclenumbermatrix}
    \renewcommand{\thefigure}{1}

    \end{figure}

\subsubsection{Interaction effects $GM(i)$ between nodes}
    \label{interaction effects}
   In this part, we calculate the interaction effects between nodes, which are evaluated based on the local properties of the nodes, the effective distance between them, and their positions within the network.
   Typically, in a network, a higher node degree indicates greater influence. A shorter distance between a node and its neighbouring nodes suggests that the node occupies a central position locally, enhancing its spreading capabilities. When a node serves as a bridge between neighbouring nodes that are not directly connected, it plays a critical role in maintaining the network's connectivity. Based on the analysis above, in calculating the interaction effects between nodes, the node's influence is measured by its degree, the distance between nodes is assessed using the effective distance, and the network constraint coefficient is applied to emphasize the importance of bridging nodes. Finally, the gravity model is employed to aggregate the information from neighbouring nodes.
     The interaction effects between nodes can be calculated as follows.
\begin{equation}
\label{gravitymodel}
    GM(i) =\sum_{ED_{(j|i)}\leq R,j\neq i}^{} e^{-c(i)}\frac{k_i \times k_j}{ED_{j|i}^2}
\end{equation}
where $R$ is the truncation radius, $c(i)$ is the network constraint coefficient, $ED_{j|i}$ is the effective distance from node $i$ to node $j$, and $k_i$ represents the degree of node $i$.
Take node 5 in the example network in Figure. \ref{framework} as an example, Figure. \ref{cyclenumbermatrix}(c) shows how the gravity model aggregates
neighbor information.

\subsubsection{Information propagation effects of nodes $RCP(i)$ }
\label{information propagation}
In this part, we evaluate the information propagation effects of nodes by quantifying their involvement in the cycles of neighbouring nodes using the cycle ratio. Taking the node's cycle ratio as the initial propagation information, we can efficiently obtain the cycle structure of the node's higher-order neighbourhood within just a few iterations. The restricted cycle ratio propagation of node $i$ is defined as follows.

\begin{equation}
    \label{rcp1}
    I_0(i)=CR(i)\quad
    I_l(i)=\frac{1}{l^2} \sum_{j\in \Lambda _i}^{} I_{l-1}(j)
\end{equation}
\begin{equation}
    \label{rcp2}
    RCP(i) =\sum_{l=1}^{T} I_l(i)
\end{equation}
where $T$ is the number of iterations, and $\Lambda _i$ is the set of neighbors of the node $i$. Taking node 4 in Figure. \ref{framework} as an example, the neighbours of node 4 within two orders are 3, 5, 6, and 7. The CR values for nodes from 3 to 7 are 0, 0, 2, 3.5, and 3.5, respectively. 
As shown in Figure. \ref{cyclenumbermatrix}(d), we calculate the first-order and second-order information propagation of node 4: $I_0(4) = CR(4)$, $I_1(4) = \frac{I_0(5)}{1^2}=\frac{CR_5}{1^2}$, and $I_2(4)=\frac{I_1(5)}{2^2}$, where $I_1(5)=\frac{CR_3+CR_6+CR_7}{1^2}$.
Finally, $RCP_4 =  I_0(4)+I_1(4)+I_2(4) = CR(4) +\frac{CR_5}{1^2} +\frac{CR_3+CR_6+CR_7}{2^2} =3.75$.

The value of the cycle ratio for each node can be obtained through Eq. \ref{cr1}.

\begin{equation}
\label{cr1}
\mathrm{CR}{(i)}=\left\{\begin{array}{ll}0, & c_{i i}=0 \\\sum_{j, c_{i j}>0} \frac{c_{i j}}{c_{j j}}, & c_{i i}>0\end{array}\right\} 
   \end{equation}
where $c_{ii}$ represents the number of cycles associated with node $i$, and $c_{ij}$ represents  the number of cycles that
pass through $i$ and $j$.  In the example network shown in Figure. \ref{framework}, all shortest cycles [0,1,2], [2,6,7], [5,6,7]. Taking node 6 as an
example, the cycle number matrix is shown in Figure. \ref{cyclenumbermatrix}(a), there are two cycles associated with node 6, its neighbouring nodes are 2, 5, and 7, so $CR_6 = \frac{1}{2} +\frac{1}{1}+\frac{2}{2}+\frac{2}{2}=3.5$.

\subsubsection{Weighted fusion}
 \label{weighted fusion}
We introduce a balancing factor $\gamma$, defined as the ratio of $GM$ to $RCP$. This adjustment is necessary because of the disparity in scales between $GM$ and $RCP$, they cannot be integrated directly.
\begin{equation}
\label{factor}
    \gamma = \frac{\left \langle GM \right \rangle }{\left \langle RCP \right \rangle }
\end{equation}
where  $\left \langle GM \right \rangle$ and $\left \langle RCP \right \rangle$ denote the average values of $GM$ and $RCP$, respectively.

  In summary, the interaction effects between nodes can be calculated by the gravity model with the effective distance of nodes and structural holes, which considers the network's global and local information. Moreover, the distance between a node and its neighbours can be made more variable by introducing the effective distance, which is consistent with the law of information propagation. Additionally, by incorporating structural holes into our methodology, HGC can more accurately identify nodes that are not central but important. We use cycle ratio to quantify the importance of cycle structure within networks. Then, the importance of a node can be evaluated by measuring the extent to which it is involved in the associated shortest cycles of other nodes. Furthermore, we can obtain the cycle structure information of high-order neighbours with just a few iterations.
  The pseudo-code of  HGC is presented in Algorithm \ref{algorithm}.

\begin{algorithm}

  \SetKwData{Left}{left}\SetKwData{This}{this}\SetKwData{Up}{up}
  \SetKwFunction{Union}{Union}\SetKwFunction{FindCompress}{FindCompress}
  \SetKwInOut{Input}{input}\SetKwInOut{Output}{output}

  \Input{graph: $G=<V,E>$ ,truncation radius $R=2$, iterations $T=2$}
  \Output{$Rank[v,HGC(v)]$}
  \BlankLine
 
  \For{$i\leftarrow 1$ \KwTo $N$}{
        Compute the degree of node $v_i$ \\

      Compute effective distance between node $v_i$ and it's neighbors using Eq. \ref{effdistance} \\
        
         Compute structural hole of node $v_i$ using Eq. \ref{strhole}\\
        
        Compute cycle ratio of node $v_i$ using Eq. \ref{cr1}\\
         Find all neighbors $\Gamma(v_i)$ of node $v_i$ within the truncation radius $R$

          }
\For{$i\leftarrow 1$ \KwTo $N$}{
 
\For{node  in $\Gamma(v_i)$}{

 Compute $GM_(v_i)$ of node $v_i$ using Eq. \ref{gravitymodel}\\
Compute $RCP(v_i)$ of node $v_i$ using Eq. \ref{rcp2}\\

 Compute the balancing factor $\gamma$ using Eq. \ref{factor}\\
}

}

\For{$i\leftarrow 1$ \KwTo $N$}{
Compute $HGC(v_i)$ of node using $v_i$ using Eq. \ref{hgm}\\
}

return $Rank[v,HGC(v)]$
  \caption{ The proposed method HGC}\label{pseudocode}
  \label{algorithm}
\end{algorithm}

\section{Experiments setup}
In this section, we describe the experimental setup used to validate the effectiveness of our proposed method HGC. We begin by introducing the datasets utilized in our experiments (Section \ref{datasets}), followed by a discussion of the methods selected for comparison (Section \ref{methods to compare}). Next, we outline the implementation of the SIR model, which serves as the foundation for our simulations (Section \ref{sir}). Finally, we present the evaluation indicators used to assess the performance of the methods (Section \ref{evaluation}).

\label{section4}
\subsection{Datasets}
\label{datasets}
	The effectiveness of the proposed method is evaluated in this paper by analysing eight real-world networks from six categories. These networks include a transportation network USAir, a communication network Email, a biology network Yeast, an infrastructure network Power, two collaborative networks including Jazz and NS, and two social networks including Facebook and PB. 
 Table \ref{network_stats} lists the basic topological information of these networks. $N$ represents the number of nodes in the network, $M$ represents the number of edges, $<k>$ represents the average degree of the nodes, $<d>$ represents the average distance between nodes, and $C$ represents the clustering coefficient of the network.
 All of the data is available and can be downloaded from \url{https://github.com/MLIF/Network-Data}.

\begin{table}[h]
\setlength{\abovecaptionskip}{0cm} 
\setlength{\belowcaptionskip}{0cm}
\centering
\caption{The topological features of eight real-world networks.}
\label{network_stats}
\renewcommand\arraystretch{1.2}
\setlength{\tabcolsep}{30pt}
\resizebox{1.0\linewidth}{!}{

\begin{tabular}{cccccc}
    
  \toprule
    Networks & \(N\) & \(M\) & \(<k>\) & \(<d>\) & \(C\)\\
  \midrule
  Jazz	& 198 &	2742	& 27.6970&	2.2350&	0.6334\\
  USAir	& 332&	2126	& 12.8072	& 2.7381& 	0.7494\\
  NS	& 379& 	914	& 4.8232& 	6.0419	& 0.7981\\

Email& 	1133	& 5451& 	9.6222& 	3.6060& 	0.2540\\
PB	& 1222	& 16714	& 27.3552	& 2.7375& 0.3600\\
Yeast	& 2361& 	7182	& 5.6289	& 3.8832& 	0.1301\\
Facebook& 	4039	& 88234& 	43.6910& 	3.6925	& 0.6170\\
Power	& 4941	& 6594	& 2.6691& 	18.9892	& 0.1065\\

  \bottomrule
  
\end{tabular}
}
\end{table}

\subsection{Methods to compare}
\label{methods to compare}
We selected several traditional and relevant methods to compare the proposed HGC method's performance comprehensively. Below is a brief introduction to these methods.
In this paper, we set $R=2$ and $T=2$ due to the small-world property \cite{smallworld} in most real-world networks.

\begin{itemize}
\item Degree centrality \cite{DC} (DC) is a measure to determine the importance of a node within a network. It is based on the number of direct connections that a node has with other nodes in the network.
\item Betweenness centrality \cite{bc} (BC) measures the extent to which a node lies on the shortest paths between other nodes.
\item Closeness centrality \cite{cc} (CC) is a measure to indicate how close a node is to all other nodes in a network.
\item The k-shell decomposition method \cite{ks} (KS) is used to analyze the structure properties of nodes and edges in a network. Initially, starting from $k=1$, remove all nodes and their edges of degree 1, these nodes belong to the $1-shell$. In the remaining graph, the process is repeated: all nodes with a degree less than $k$ and their edges are removed, these nodes belong to the $k-shell$. Finally, the decomposition process continues until no more nodes can be removed.	
\item Cycle ratio \cite{cr} (CR) can be calculated to quantify the importance of an individual node by simply measuring to which extent it is
involved in other nodes’ associated shortest cycles.
\item The local gravity model \cite{GM} (LGM) uses both the neighbourhood and path information. Furthermore, it can reduce the computational complexity by truncation radius.
\item The restricted degree information propagation \cite{RDP} (RDP) can obtain the degree information of higher-order neighbours
by only considering the directly connected neighbours.
\end{itemize}

\subsection{SIR model}
\label{sir}
\begin{figure*}[htpb]
    \centering
    \includegraphics[width=1.0\linewidth]{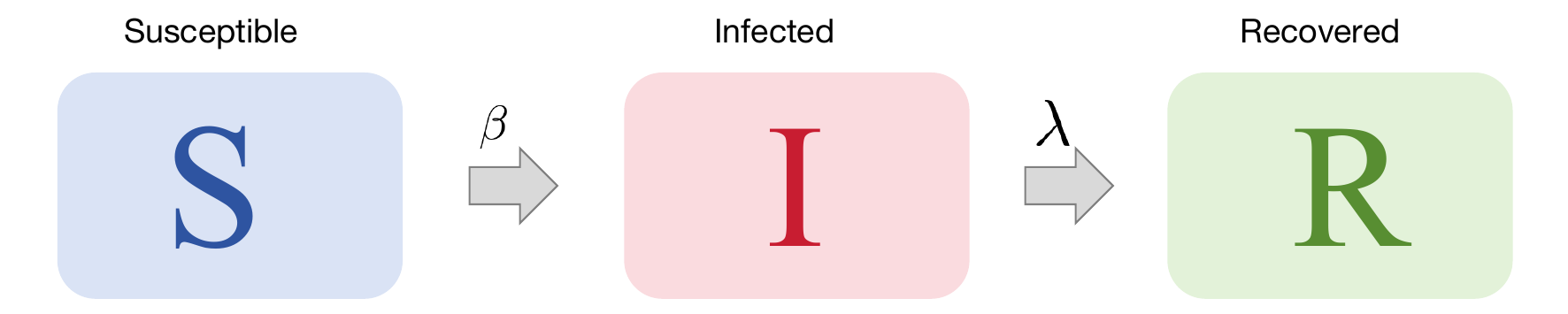}
    \caption{ Schematic diagram of SIR model state transition.
    }
    \label{SIR}
    \renewcommand{\thefigure}{1}

    \end{figure*}

The SIR model \cite{sir} is a classic epidemic model used to describe the spread of infectious diseases within a population. The SIR model assumes that nodes have three states: Susceptible (S), Infected (I), and Recovered (R). Initially, a single node is randomly selected to be in the I state, and the other nodes are in the S state. At each time step, the infected nodes will infect their susceptible neighbours with probability $\beta$, and then each infected node changes to the R state with probability $\lambda=1$. Figure. \ref{SIR} is the schematic diagram of SIR model state transition.
The process continues until there are no nodes in the I state. Then, $F(t)$ represents the total number of nodes in states $I$ and $R$ at time $t$. As $t$ increases, $F(t)$ also increases and eventually approaches a steady state. To minimize the interference of random factors, values of propagation influence for networks are the averages of 1000 independent experiments.
Furthermore, the epidemic threshold \cite{threshhold} of a network is expressed as $\beta _c\approx \frac{\left \langle k \right \rangle }{\left \langle k^2 \right \rangle -\left \langle k \right \rangle } $, where $\left \langle k \right \rangle$ and $\left \langle k^2 \right \rangle$ are the average degree and the second-order moment of the degree distribution.

\subsection{Evaluation indicators}
\label{evaluation}

The Kendall’s Tau \cite{kendall}  is a statistical measure used to assess the correlation between two ranked sequences. It reflects the degree of association between two variables by calculating the consistency of their orderings. If in both sequences, the element ranked higher/lower in the first sequence is also ranked higher/lower in the second sequence,  it is called a concordant pair. Otherwise, it is called a discordant pair. The Kendall's Tau of two sequences is defined as
\begin{equation}
    \tau = \frac{2(n_c-n_d)}{N(N-1)}
\end{equation}
where $n_c$ and $n_d$ are the numbers of concordant pairs and discordant pairs. The higher the value of $\tau$, the more accurate the ranked list obtained by this method.

In certain scenarios, ranking all nodes in the network is not a primary concern; instead, we only need to identify the top-k important nodes, which is particularly crucial for controlling information dissemination. Therefore, in addition to evaluating the ranking of all nodes in the network, we also assess the performance of these methods in identifying the top-k influential spreaders in the network. In other words, we compare the descending order lists of nodes obtained from the method with those obtained from SIR simulations. We focus only on the top-k nodes in each list to analyze the similarity between these two lists. The Jaccard similarity coefficient \cite{jaccard} is used to assess the similarity between the top-k nodes in two ranking lists. The Jaccard similarity coefficient of two sequences is defined as
\begin{equation}
    Jaccard(X,Y)=\frac{\left | X\cap Y \right | }{\left |  X\cup Y\right  | }
\end{equation}
where $X$ and $Y$ represent the top-k nodes identified by two methods. The greater the coefficient value, the greater the degree of similarity between two ranked lists.

The monotonicity \cite{mono} is used to measure the ability of different methods to distinguish the importance of nodes, and it is defined as
\begin{equation}
    M(L) = [1-\frac{ {\textstyle \sum_{r\in L}N_r(N_r-1)^{}} }{N(N-1)}]^2
\end{equation}
where $N$ is the size of list $L$, and $N_r$ is the number of nodes with same rank. The value of $M$ ranges from [0,1]. When $M=1$, it means that each rank in the ranked list $L$ contains a single node, indicating that the current ranking method exhibits the best distinguishing ability. In contrast, $M=0$  indicates that the current ranking method has the poorest distinguishing ability.

\section{Results and discussion}

\label{section5}

In this section, we conducted several experiments on eight datasets of different sizes, evaluating them from accuracy, spreading ability, and distinguishing ability.

\subsection{The comparison of accuracy}

Table \ref{kendall} presents the accuracy results measured by Kendall’s Tau of different methods across eight networks.
 According to Table \ref{kendall}, our method HGC achieved the best performance in Jazz, NS, Email, PB, and Facebook networks due to its incorporation of cycle structure and effective distance, which are more prominent in these densely connected networks. In the USAir and Yeast networks, HGC shows minimal difference compared to the top-ranked method. However, in the Power network, due to its sparse structure with relatively few connections between nodes, the cycle structure information in the network does not play a significant role. Despite this, HGC achieved the second-best performance among the seven methods evaluated.

\begin{table}[htbp]
\setlength{\abovecaptionskip}{0cm} 
\setlength{\belowcaptionskip}{0cm}
\centering
\caption{The methods’ accuracies for \(\beta = \beta_c\) , measured by the Kendall’s Tau. The top-ranked value in
each row of the table is marked in bold, while the second-ranked value is underlined.}
\label{kendall}
\renewcommand\arraystretch{1.2}
\setlength{\tabcolsep}{15pt}
\resizebox{1.0\linewidth}{!}{
\begin{tabular}{ccccccccc}
    
  \toprule
  Networks & DC & BC & CC & KS & CR&LGM&RDP&HGC\\
  \midrule
 Jazz	&0.8145&	0.5008	&0.7113&	0.7465	&0.6327&	\underline{0.855}&	0.8326&	\textbf{0.865}\\
NS	&0.63	&0.3559&	0.3248&	0.5606	&0.3699&	\underline{0.8003}&	0.764	&\textbf{0.8107}\\
USAir&	0.7378&	0.5577&	0.7098	&0.7324&	0.4505&	\textbf{0.7758}&	0.7256&	\underline{0.766}\\
Email&	0.7933	&0.6786&	0.7631&	0.8035	&0.5867&	\underline{0.8485}&	0.8221&	\textbf{0.8609}\\
Yeast	&0.6655&	0.6259	&0.723	&0.7356	&0.6017	& \textbf{0.8099}&	0.7656&	\underline{0.8085}\\
Power&0.6066&	0.4697	&0.2295&	0.4098&	0.3462	&\textbf{0.7383}	& 0.7007&	\underline{0.7145}\\
PB	&0.8223	&0.6776&	0.7298	&0.8087&	0.6399&	\underline{0.8277}&	0.8145	&\textbf{0.8388}\\
Facebook	&0.6798&	0.5156	&0.3338	&0.7075&	0.4022	&\underline{0.7473} &	0.6876&	\textbf{0.7575}\\
  \bottomrule
\end{tabular}
}
\end{table}

\begin{figure*}[htbp]
    \centering
    \includegraphics[width=1.0\linewidth]{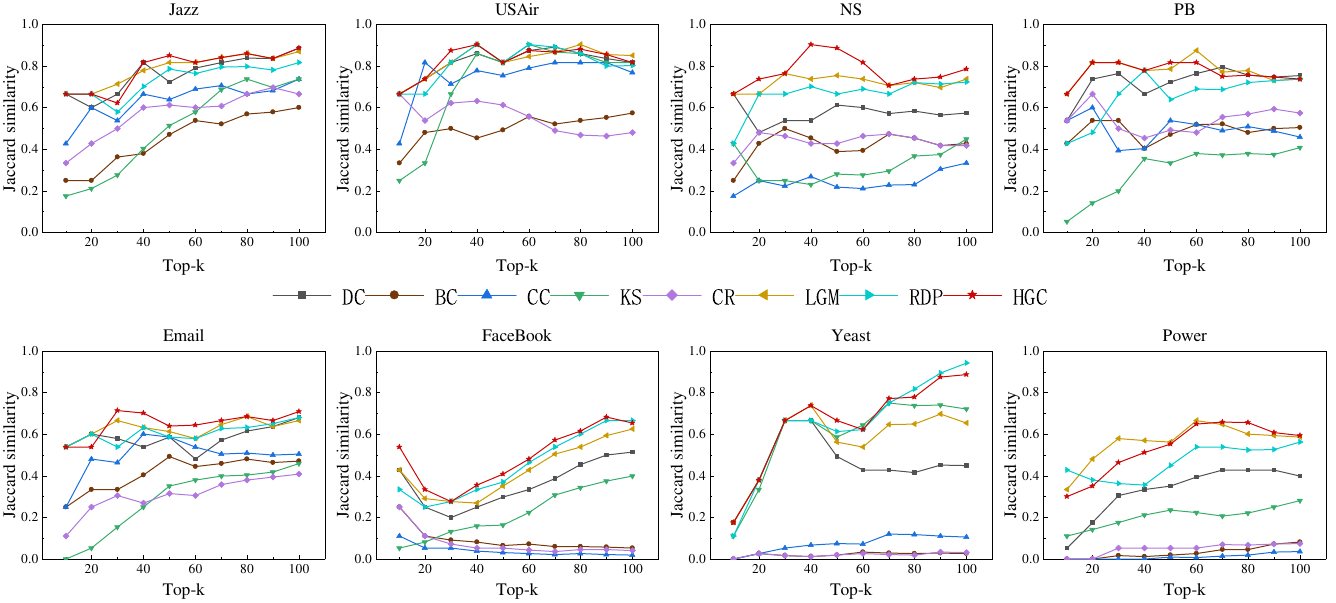}
    \caption{ The Jaccard similarity coefficients on the top-k influential spreaders.  The X-axis represents the number of top influential spreaders, and the Y-axis represents the value of the Jaccard similarity coefficient.}
    \label{graph_jaccard}
    \renewcommand{\thefigure}{1}

    \end{figure*} 
\begin{figure*}[htbp]
    \centering
    \includegraphics[width=1.0\linewidth]{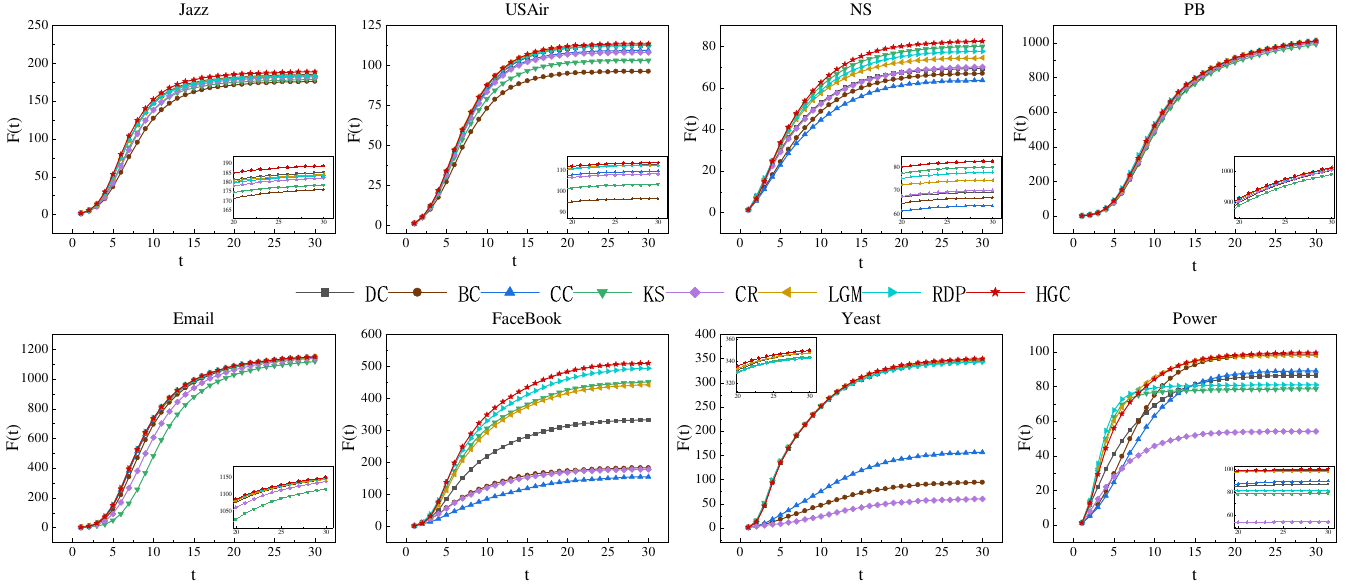}
    \caption{ The propagation influence of the top-10 ranking nodes of various methods simulated by SIR (under $\beta=\beta_c$). The X-axis represents the time step, and the Y-axis represents the number of infected and recovered nodes at time $t$. }  
    \label{graph_sir}
    \renewcommand{\thefigure}{1}

    \end{figure*} 

\begin{table}[h]
\setlength{\abovecaptionskip}{0cm} 
\setlength{\belowcaptionskip}{0cm}
\centering

\caption{Monotonicity of node-ranked sequences derived by the eight methods. The top-ranked value in
each row of the table is marked in bold, while the second-ranked value is underlined.}
\label{Monotonicity}

\renewcommand\arraystretch{1.2}
\setlength{\tabcolsep}{15pt}
\resizebox{1.0\linewidth}{!}{
\begin{tabular}{ccccccccc}
    
  \toprule
  Networks & DC & BC & CC & KS & CR&LGM&RDP&HGC\\
  \midrule
Jazz    &	0.9659&	0.9885&	0.9878	&0.7944	&0.9985&	\underline{0.9993}&	0.9992&	\textbf{0.9994}\\
NS	&0.7642	&0.339	&0.9928	&0.6421	&0.983	&\underline{0.9949}&	0.9933&	\textbf{0.9952}\\
USAir&	0.8586&	0.697&	0.9892	&0.8114&	0.9451&	\textbf{0.9951}&	\underline{0.995}	&\textbf{0.9951}\\
Email&	0.8874&	0.94	&0.9988	&0.8088	&0.9631&	\textbf{0.9999}&	\underline{0.9998}	&\textbf{0.9999}\\
Yeast&	0.8314&	0.8293	&0.9988	&0.7737&	0.8831&	\underline{0.9991}&	0.9986&	\textbf{0.9992}\\
Power&	0.5927&	0.8319	&\underline{0.9998}	&0.246	&0.7653	&0.9984&	0.9903&	\textbf{0.9999}\\
PB	&0.9328	&0.9489&	\underline{0.998}&	0.9064	&0.9748&	\textbf{0.9993}&	\textbf{0.9993} & \textbf{0.9993}\\
Facebook&	0.9739&	0.9855&	0.9967	&0.9419	&\underline{0.9993}	&\textbf{0.9999}	&\textbf{0.9999}	&\textbf{0.9999}\\
  \bottomrule
\end{tabular}
}
\end{table}

Figure. \ref{graph_jaccard} illustrates the results of the Jaccard similarity coefficient for identifying the top-k influential spreaders, ranging from 10 to 100, with a step size of 10. According to Figure. \ref{graph_jaccard}, we can observe that our method HGC performs best in the NS and Facebook networks, achieving the highest Jaccard similarity coefficients, indicating its superior capability in identifying important nodes. In the Jazz, Email, and Power networks, the initial performance of HGC is suboptimal. However, as we increase the number of top-k nodes being considered, the similarity improves, suggesting that HGC becomes more effective when evaluating a larger portion of the network. In the USAir, PB, and Yeast networks, HGC remains competitive, performing comparably or slightly better than other methods. Therefore, 
it can be concluded that HGC not only effectively ranks all nodes in the network but also correctly identifies the top-k most important nodes.

\subsection{The comparison of spreading ability}

 Figure. \ref{graph_sir}  shows the propagation influence of the top 10 nodes obtained by various methods across eight networks, as well as the influence between time 20 and 30 to better highlight the differences.
 In the Jazz, USAir, NS, and Facebook networks, our method  HGC demonstrates superior infection performance, significantly outperforming other methods.
 In PB networks, the dense nature of the connections allows infected nodes to quickly infect their neighbours, making it easier for all methods to perform well. As a result, the variance in performance between methods is reduced.
 HGC performs similarly to RDP and LGM in the Yeast and Email networks, significantly better than traditional methods like BC and CC. Our method initially did not perform the best in the power network, but its performance improved over time. This improvement is attributed to HGC's consideration of the cycle structure information within the network, leading to the identification of important nodes that are heavily involved in cycles and thus possess better propagation ability. In conclusion, the important nodes identified by HGC exhibit excellent propagation performance.

\subsection{The comparison of distinguishing ability}

Table \ref{Monotonicity} presents the monotonicity results of different methods across eight networks. Our proposed method
HGC achieves the highest monotonicity values in all eight networks, demonstrating its strong capability to identify important nodes and uncover deeper layers of node information. Additionally, the RDP method achieved the best performance in the Email network, while both RDP and LGM methods also attained the highest monotonic values in the PB and Facebook networks. Overall, HGC outperforms all comparison methods in distinguishing the importance of nodes.

\section{Conclusion}
\label{section6}

This paper proposes a hybrid method that combines the gravity model and cycle structure to identify influential spreaders in networks. 
The gravity model is used to calculate the interaction influence between adjacent nodes, and effective distance is used to more accurately measure the distance between nodes, addressing the issue of symmetrical influence in the gravity model. Additionally, we utilize the network constraint coefficient to quantify structural holes, enabling the identification of non-central but important nodes.
Furthermore, the cycle structure information of nodes in the network is incorporated to evaluate their influence comprehensively.
To validate the effectiveness of the proposed method, eight real-world networks with different structural attributes were selected, seven methods were compared, and experiments were conducted from three aspects: the accuracy of node identification, the spreading capability of nodes, and the ability to distinguish nodes.
From the experimental results, the proposed method HGC demonstrates superior performance in identifying important nodes within various network structures. 
However, given the small-world property of most real-world networks, HGC is currently limited to considering neighbouring nodes up to the second order. Determining the appropriate truncation radius for networks with varying structures remains an open challenge.



\bibliographystyle{elsarticle-num} 
\bibliography{ref}






\end{document}